\documentclass{aa}  

\usepackage{graphicx}
\usepackage{txfonts}
\usepackage{braket}
\usepackage{mathtools}
\usepackage{hyperref}

\begin{document}

\title{Graph-based clustering of gamma-ray bursts}

\author{Mariusz Tarnopolski\inst{1}}

\institute{Astronomical Observatory, Jagiellonian University, Orla 171, 30-244 Krak\'ow, Poland (\email{mariusz.tarnopolski@uj.edu.pl})}

\date{Received DATA accepted }

\abstract
{}
{ An attempt to classify gamma-ray bursts (GRBs) with a low level of supervision using the state-of-the-start approaches stemming from graph theory was undertaken. }
{ Graph-based classification methods, relying on different variants of the $k$-nearest neighbour graph, were applied to various GRB samples in the duration--hardness ratio parameter space to infer the optimal partitioning.}
{ In most cases it is found that both two and three groups are feasible, with the outcome being more ambiguous with an increasing sample size. }
{ There is no clear indication of the presence of a third GRB class; however, such a possibility cannot be ruled out with the employed methodology. There are no hints at more than three classes though.}


\maketitle

\section{Introduction}
\label{introduction}

Gamma-ray bursts (GRBs, \citealt{klebesadel}) are confidently divided into two classes: short (attributed to compact-object mergers) and long (massive-star collapsars). The dichotomy is apparent in the bimodal distribution of durations $T_{90}$ (i.e. the time during which 90\% of the GRB's fluence is detected), and it occurs at $T_{90}\simeq 2\,{\rm s}$ (\citealt{kouve93}). However, this is not a sharp separation due to significant overlap (\citealt{bromberg13,tarnopolski15a}; see also \citealt{ahumada21} for the shortest confirmed GRB from a collapsar). Since the third, intermediate-duration class was reported \citep{horvath98}, the distribution has been routinely modelled with a mixture of normal distributions in several subsequent works (e.g. \citealt{horvath02,horvath08,zhang08,huja09,horvath10,zhang16}), which have often concluded that a third Gaussian component is required to fit the data appropriately and have attributed physical meaning to it.

However, this third component is not necessarily evidence of a physically motivated group. Indeed, it is likely a signature of inherent skewness of the long GRB group \citep{koen,tarnopolski15b,tarnopolski20}. It follows that when modelling with skewed distributions, instead of symmetric ones (e.g. Gaussian or Student), only two components are required to model the data appropriately \citep{tarnopolski16a,tarnopolski16c,kwong}, implying the third one is spurious, and it appears because of modelling an intrinsically skewed distribution with symmetric ones.

Investigating the two-dimensional realm of hardness ratios and durations has led to analogous conclusions: Gaussian mixtures have often pointed at three groups \citep{horvath06,ripa09,horvath10,veres10,horvath18}, but several works have indicated that only two are required \citep{ripa12,yang16,tarnopolski19d,kienlin20}. Moreover, considering skewed distributions, two-component mixtures have also been indicated \citep{tarnopolski19d,tarnopolski19a}. In higher-dimensional spaces, things become less unambiguous \citep{mukh,chattopadhyay07,chattopadhyay17,chattopadhyay18,modak18,acuner,horvath19,toth19,tarnopolski19b}. For the most recent, detailed overview on the topic of parametric clustering of GRBs, readers can refer to \citet{tarnopolski19b}.

A non-parametric approach to determining the number of classes is a desirable route that has been undertaken on a few occasions. \citet{mukh} performed average linkage hierarchical agglomerative clustering, which however yielded ambiguous results, pointing at either two or three groups. \citet{balastegui} claimed the existence of a third class based on neural network classification; however, \citet{hakkila00,hakkila03} attributed the presence of this class to instrumental effects and questioned its physical reality. This conclusion was also supported by \citet{rajaniemi}, who employed an independent analysis method (self-organising map; \citealt{kohonen}). In particular, the outputs of such unsupervised classifications are affected by various factors, for example, the utilised technique, the specificity of the samples and attributes used, among others \citep{hakkila04}, and also by systematic biases \citep{roiger00}. \citet{chattopadhyay07}, on the other hand, used different clustering methods ($K$-means and Dirichlet process; the latter with an underlying assumption of a multi-normal distribution), and again found statistical evidence for three GRB classes. Based on the $K$-means method as well, \citet{veres10} claimed evidence for the third class. The same approach turned out to be inconclusive for the Reuven Ramaty High Energy Solar Spectroscopic Imager (RHESSI) data \citep{ripa12}. However, a classification based only on the prompt light curves in different energy bands has unambiguously separated GRBs into just two classes \citep{jespersen20}.

Graph theory has been rarely used for clustering astronomical objects based on their multi-dimensional properties \citep[e.g. ][]{farrah09,maritz17}, except for spatial clustering on the celestial sphere \citep{campana08,tramacere13}, or the detection of filamentary structures \citep{bonnaire20}. With the often contradicting results attained for GRBs so far, it is a path worth exploring.

This paper is structured as follows. Section~\ref{data} characterises the GRB samples. A brief introduction to the basic concepts of graph theory is provided in Sect.~\ref{theory}. In Sect.~\ref{methods} the employed graph-based clustering methods are described. The results are presented in Sect.~\ref{results}. They are discussed in Sect.~\ref{discussion}, followed by concluding remarks gathered in Sect.~\ref{summary}. The \textsc{Matlab}~2018a and \textsc{Mathematica}~v12.0 computer algebra systems are utilised throughout.

\section{Data}
\label{data}

The Fermi data set from the third catalogue \citep{bhat} consists of 1376 GRBs with measured $T_{90}$ and $H_{32}$, where the hardness ratio $H_{32}=\frac{F_{50-300\,{\rm keV}}}{F_{10-50\,{\rm keV}}}$ is the ratio of fluences $F$ in the respective energy bands during the $T_{90}$ interval. This sample was investigated previously using mixture models \citep{tarnopolski19a}. Herein, four outliers with $\log H_{32}>1.4$ are excluded, resulting in 1372 GRBs. The Burst And Transient Source Experiment\footnote{\url{https://gammaray.nsstc.nasa.gov/batse/grb/catalog/current/}} (BATSE) onboard the Compton Gamma-Ray Observatory observed 1954 GRBs with $T_{90}$ and $H_{32}$, with the hardness ratio computed within slightly different energy bands: $H_{32}=\frac{F_{100-300\,{\rm keV}}}{F_{50-100\,{\rm keV}}}$. This sample was also analysed by \citet{tarnopolski19a}. Herein, 1953 GRBs are utilised as one point has both $\log T_{90}$ and $\log H_{32}$ below $-1$, and it is excluded as an obvious outlier.

The following four data sets are also considered: 1028 GRBs from the Swift Burst Alert Telescope catalogue \citep{lien16}, 1143 GRBs observed by Konus-Wind \citep{svinkin16}, 426 GRBs detected by RHESSI \citep{ripa09}, and 257 GRBs from Suzaku Wide-Band All-Sky Monitor \citep{ohmori}. For each instrument, fluences in different energy bands are available, hence the definitions of $H_{32}$ are as follows: $H_{32}=\frac{F_{50-100\,{\rm keV}}}{F_{25-50\,{\rm keV}}}$ for Swift; $H_{32}=\frac{F_{200-750\,{\rm keV}}}{F_{50-200\,{\rm keV}}}$ for Konus; $H_{32}=\frac{F_{120-1500\,{\rm keV}}}{F_{25-120\,{\rm keV}}}$ for RHESSI; and $H_{32}=\frac{F_{240-520\,{\rm keV}}}{F_{110-240\,{\rm keV}}}$ for Suzaku. These were examined with mixture models as well \citep{tarnopolski19d}; compared to this previous work, herein two duplicates in the Suzaku sample were removed, one duplicate in the case of RHESSI, and three outliers (with $\log H_{32}<-0.3$) and two duplicates from Swift.

Focus is laid on the two-dimensional parameter space spanned by $T_{90}$ and the appropriate $H_{32}$. The reasons are that (i) it has an advantage of being easily displayed graphically, consequently being easy to comprehend at a glance, and (ii) it has been widely explored in the literature allowing for straightforward comparisons. Adding other quantities, for example, fluence \citep[][and references therein]{tarnopolski19b} to form higher-dimensional spaces is left for a future study.

\section{Graph theory}
\label{theory}

\begin{figure}
\centering
\includegraphics[width=0.7\columnwidth]{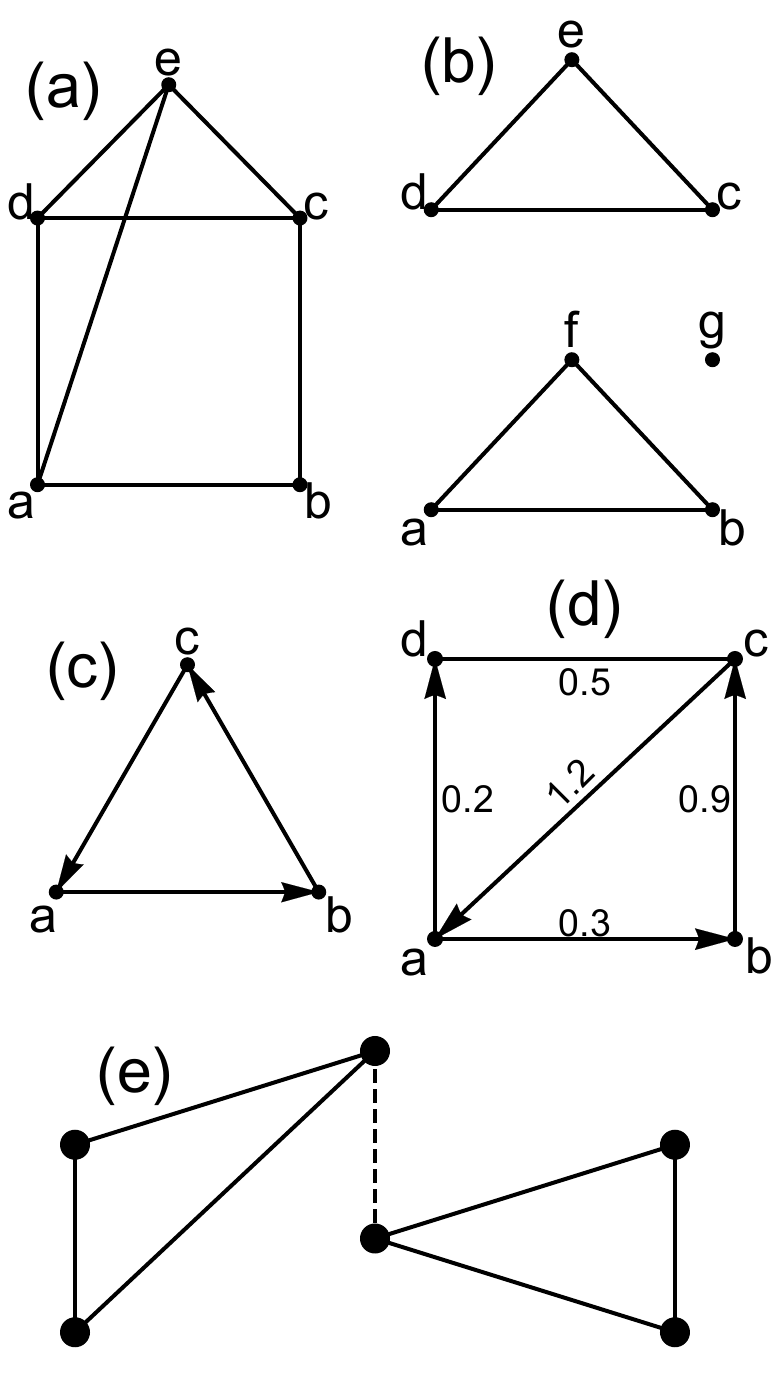}
\caption{Exemplary graphs. (a) An undirected, connected planar graph -- the edge $ae$ can be drawn to the left of vertex $d$, so that it will not intersect with the edge $cd$. Edges need not be straight lines. (b) A disconnected graph. Degree-zero vertices, such as $g$ here, are allowed. (c) A directed graph. A path from, e.g. $b$ to $c$ is allowed, but to go back one needs to travel from $c\to a\to b$. (d) A mixed weighted graph. (e) The dashed edge is a bridge. After removing this edge, the graph becomes disconnected, so it yields $\lambda (G)=1$. }
\label{fig1}
\end{figure}

A graph \citep{wilson} $G=(V,E)$ is a collection of vertices $V$ connected by edges $E$. Two vertices need not be connected directly, for example, $a$ and $c$ in Fig.~\ref{fig1}(a). An edge can be considered as a binary operator such that $\delta_{ij}=1$ if $i,j$ are adjacent vertices, and $\delta_{ij}=0$ otherwise. If every vertex is joined by an edge with every other vertex, the graph is called a complete one. A series (of any length) of edges connecting two vertices forms a path, for example, $a\to b\to c$. If any two vertices in $G$ can be joined by a path, then $G$ is called a connected graph. A disconnected graph consists of two or more components [Fig.~\ref{fig1}(b)]. In a simple graph, two vertices are joined by one edge at most. A simple graph has no loops, that is, no edges connecting a vertex to itself. If there are multiple edges joining two vertices one deals with a multigraph. Edges terminate in vertices. Intersecting edges do not create a new vertex. The degree $\deg(v)$ of a vertex $v$ is the number of edges incident to this vertex. A graph that can be drawn in a plane so that no edges intersect is called a planar graph. In the case of graphs, the relative orientation of vertices does not play a role. Hence, by rearranging the vertices of $G$ in a way that the same edges connect the same vertices, one obtains a graph $G'$ that is isomorphic with $G$. 

Let $u$ and $v$ be adjacent vertices, so $e=uv$ is an edge. It is undirected if one can travel directly from $u$ to $v$ and vice versa. It is directed if only one direction is allowed [Fig.~\ref{fig1}(c)]. An undirected graph is one with all of its edges being undirected. Similarly, a directed graph has all of its edges directed. A mixed graph has both directed and undirected edges. A weighted graph has some weights associated to its edges [regardless of whether they are directed or undirected; Fig.~\ref{fig1}(d)].

If a connected graph can be disconnected by removing some particular $n$ edges, but not any $n-1$ edges, its edge connectivity is $\lambda(G)=n$. In particular, if a graph contains a single edge whose removal disconnects the graph, this edge is called a bridge [Fig.~\ref{fig1}(e)]. Likewise, a graph's vertex connectivity $\kappa(G)$ is the minimal number of vertices whose removal results in a disconnected graph.

For an undirected graph, the handshaking theorem holds:
\begin{equation*}
\sum\limits_{v\in V}\deg(v) = 2|E|,
\end{equation*}
where $|E|$ is the total number of edges. A graph can be summarised by its adjacency matrix $A$, whose elements are $A_{ij}=1$ if vertices $i$ and $j$ are connected by an edge, and $A_{ij}=0$ otherwise. For instance, for the graph in Fig.~\ref{fig1}(a),
\begin{equation*}
A=
\begin{pmatrix}
 0 & 1 & 0 & 1 & 1 \\
 1 & 0 & 1 & 0 & 0 \\
 0 & 1 & 0 & 1 & 1 \\
 1 & 0 & 1 & 0 & 1 \\
 1 & 0 & 1 & 1 & 0
\end{pmatrix}
,
\end{equation*}
where $A_{12}=1$ means that vertices $a$ and $b$ are joined by an edge, but since $A_{13}=0$ there is no edge between $a$ and $c$, and so on. For a simple graph, the diagonal is always composed of zeros since there are no loops. It follows that
\begin{equation*}
|E| = \sum\limits_{i,j}A_{ij}/2.
\end{equation*}
A degree vector $\vec{d}$ has elements $d_i\equiv\deg({i})=\sum_j A_{ij}$, and the diagonal degree matrix $D$ is defined by $D_{ii}=d_i$.

\section{Methods}
\label{methods}

In graph-based clustering methods \citep{fortunato09}, the aim is to construct a graph from multivariate data, and then partition that graph into disconnected components (communities) according to some predefined rules and objectives. The components are then associated with different clusters. A vast collection of algorithms has been gathered at \url{https://github.com/benedekrozemberczki/awesome-community-detection}. Some detect a predetermined number of communities, whereas others aim to provide the means necessary to decide both the number of clusters and the best partition. A few of the latest approaches are described in the following sections, and then applied to GRB data sets from Sect.~\ref{data}.

\subsection{Continuous $k$-nearest neighbour}
\label{sect::4.1}

The central concept of several algorithms is the $k$-nearest neighbour graph (kNN). For every point in the sample, one seeks its $k$ nearest neighbours (commonly employing a Euclidean distance measure) and connects it to them. Such an approach allows one to capture local densities, but it is often sensitive to the value of $k$. A way of circumventing this issue is the continuous kNN graph (CkNN; \citealt{berry19}). Let the (metric) distance between points $i$ and $j$ be denoted by $d(i,j)$, and the distance from $i$ to its $k$-th nearest neighbour be denoted by $d^k(i)$. Then, $i$ and $j$ are joined by an edge if $d(i,j)<\delta\sqrt{d^k(i)d^k(j)}$, where $\delta$ is a parameter controlling the sparsity of the graph.

Markov stability is used to assess the relevant partition \citep{liu20}. Let us consider a Markov process\footnote{A Markov process is a stochastic model in which the probability of each event (the transition between states) depends only on the state of the previous event. } on a graph, with a one-step transition matrix $M=D^{-1}A$. Its stationary distribution is $\vec{\pi} = \vec{d}^T/(2|E|)$, and its autocovariance matrix is $B(t) = D/(2|E|) P(t) - \vec{\pi}^T\vec{\pi}$, where the transition matrix $P(t) = \exp\left[ -t(I-M) \right]$, with $I$ being the identity matrix.

Thence, the Markov stability $r(t,g_s)$ of a partition of the nodes $g$ into $c$ non-overlapping subsets $g_s$ is defined as
\begin{equation*}
r(t,g_s) = \sum\limits_{s=1}^c\sum\limits_{i,j\in g_s}B(t)_{ij}.
\end{equation*}
A good partition is one that attains high values of the Markov stability as a function of $t$,
\begin{equation*}
r^*(t) = \max r(t,g),
\end{equation*}
achieved for some partition $g^*$. The so-called Markov time $t$ is a parameter that controls the resolution of partitions. For small $t$, many communities are expected to emerge as local structures in the graph dominate. For larger $t$, the global features become prominent. Asymptotically, the Markov stability settles on a bipartition. A robust partition persists for long Markov times (i.e. the obtained number of communities is constant), and the dissimilarity of partitions for different $t$ and $t'$ is low (i.e. the group membership of the nodes does not change rapidly). To measure the latter, the variation of information ($VI$; \citealt{meila07}) was employed, and the matrix $VI(t,t')$ contains large blocks with small values. 

The $VI$ is based on the concepts of entropy and information. Let us consider a partition $\mathcal{C}$ of $n$ points into $K$ mutually disjoint clusters $\mathcal{C}_i$. Let $n_i>0$ be the number of points in cluster $\mathcal{C}_i$. Similarly, let $\mathcal{C'}$ be another partition into $K'$ clusters (not necessarily equal to $K$), and $n'_i$ be the size of cluster $\mathcal{C'}_i$. The probability that a randomly picked point is in cluster $\mathcal{C}_i$, given the partition $\mathcal{C}$, is $P(i)=n_i/n$. Similarly, one defines $P'(i')$. Thence, the entropy associated to the partition $\mathcal{C}$ is
\begin{equation*}
\mathcal{H}\left( \mathcal{C} \right) = -\sum\limits_{i=1}^K P(i) \log P(i),
\end{equation*}
and similarly for $\mathcal{C'}$. The entropy is zero only if there is just one cluster in a partition. Let us set $P(i,i') = \left| \mathcal{C}_i \cap \mathcal{C'}_{i'} \right|/n$ to denote the joint probability that a randomly picked point belongs to clusters $\mathcal{C}_i$ and $\mathcal{C'}_{i'}$ in the respective partitions. The mutual information between the two partitions is then defined to be
\begin{equation*}
\mathcal{I}\left(\mathcal{C},\mathcal{C'}\right) = \sum\limits_{i=1}^K \sum\limits_{i'=1}^{K'} P(i,i') \log\frac{P(i,i')}{P(i)P'(i')},
\end{equation*}
and $\mathcal{I}\left(\mathcal{C},\mathcal{C'}\right) = \mathcal{I}\left(\mathcal{C'},\mathcal{C}\right)$. Finally, the $VI$ is defined as
\begin{equation*}
\begin{split}
VI\left(\mathcal{C},\mathcal{C'}\right) &= \mathcal{H}\left( \mathcal{C} \right) + \mathcal{H}\left( \mathcal{C'} \right) - 2 \mathcal{I}\left(\mathcal{C},\mathcal{C'}\right) \\
&\equiv \mathcal{H}\left( \mathcal{C} | \mathcal{C'} \right) + \mathcal{H}\left( \mathcal{C'} | \mathcal{C} \right),
\end{split}
\end{equation*}
where $\mathcal{H}(.|.)$ is the conditional entropy. Furthermore, $VI$ is a metric and is bounded to the closed interval $\left[ 0 , \log n \right]$.

To summarise the whole clustering algorithm, one must first construct a CkNN graph, then partition it. Next, one must choose the partition that has (i) a long plateau in the number of communities against Markov time $t$ and (ii) a low $VI$ of the partitions within it. This implies that the partition is robust. A \textsc{Matlab} implementation is provided by the authors\footnote{\url{https://github.com/barahona-research-group/GraphBasedClustering}}.

\subsection{\texttt{CutPC}}
\label{sect::4.2}

The \texttt{CutPC} algorithm \citep{li20} starts by constructing a variant of the kNN graph, that is, the natural neighbour (NaN) graph (NNG; \citealt{huang16}). Two vertices $u$ and $v$ are considered to be NaNs if $u$ is a neighbour of $v$ and vice versa. For example, if one were to consider the three points 1, 3, and 4 on a real line, the nearest neighbour of 1 is 3, but the nearest neighbour of 3 is 4, hence 1 and 3 are not NaNs; however, 3 and 4 are NaNs. The definition is natural since it comes from objective reality: one should consider a friend as a person who also thinks of him or her as a friend. In general, two points are associated with each other (i.e. joined by an edge in the NNG) if they are both similar to each other (homophily). The construction of NNG requires tuning no parameters, contrary to the kNN graph, in which the number $k$ of nearest neighbours needs to be set. 

Points in sparser regions have fewer NaNs than points in denser regions, which naturally leads one to consider points in very sparse regions as outliers, or noise-induced. These can be identified with the reverse density. First, let us compute the mean NaN distance of point $i$:
\begin{equation*}
\tau (i) = \frac{1}{k} \sum\limits_{j\in \rm{NaN}(j)} d(i,j),
\end{equation*}
where $d(i,j)$ is the Euclidean distance between $i$ and $j$, and the sum runs over the $j$ that are NaNs of $i$. The reverse density is
\begin{equation*}
\theta = \braket{\tau(i)} + \alpha \sqrt{\rm{var} (\tau (i))},
\end{equation*}
where the triangular brackets denote the mean over all points in a data set, and $\alpha$ is a tuning parameter (the only free parameter in this algorithm). A point $i$ is considered an outlier if $\tau(i)>\theta$, that is, if its mean distance to NaNs exceeds $\alpha$ standard deviations from the mean. After discarding outliers and the edges connected to them from the NNG, the remaining points were divided into connected clusters. Inherently, the clusters are expected to be located around points with high density, and in principle the method should work on overlapping clusters, too, after tweaking the value of $\alpha$. The algorithm is relatively fast, with complexity of $\mathcal{O}(n\log n)$. A \textsc{Matlab} implementation is provided by the authors\footnote{\url{https://github.com/lintao6/CutPC}}.

\subsection{Graph connectivity}
\label{sect::4.4}

A connectivity-based approach to clustering \citep{li18graph} that takes cluster overlaps into account starts by constructing a kNN graph, $G^k$. As a rule of thumb, $k\geqslant 6(m-1)$, where $m$ is the dimensionality of the data. The central point of the methodology is to identify a set $S$ of {\ singular points}, whose removal will split the graph into clusters. Let us define a set of distance $d$ vertices:
\begin{equation*}
V_v(d) = \left\{ u\in V | d_{G^k}(u,v) = d\right\},
\end{equation*}
where $d_{G^k}(u,v)$ is the path length between vertices $u$ and $v$. Let us define a new graph, $G^k_{v,d,a} = \left( V_v(d), E_v(d,a) \right)$, where $a\leqslant 2d$, and the edges
\begin{equation*}
E_v(d,a) = \left\{ u_1u_2 | d_{G^k}(u_1,u_2)\leqslant a \wedge u_1,u_2\in V_v(d) \right\}.
\end{equation*}
Graph $G^k_{v,d,a}$ can be connected or disconnected, hence it is composed of one or more components, $\mathcal{C}$; their number is denoted by $\left|\mathcal{C}|G^k_{v,d,a}\right|$. The singular index $SI$ of vertex $v$ can then be approximated by the number of these components:
\begin{equation*}
SI(v|k,d,a) = \sum\limits_i \frac{1+\log|C_i \cap V_v(d)|}{\log\left(|V_v(d)|+1\right)} \approx \left|\mathcal{C}|G^k_{v,d,a}\right|,
\end{equation*}
where the sum is over all $i$ such that $C_i\in C|G^k_{v,d,a}$. A vertex $v$ is a singular point if $SI(v|k,d,a)>1$, meaning there exists more than one cluster in distance $d$ from $v$. Let us denote the graph that results from removing all singular points by $G_{-S} = (V\setminus S,E)$. The components of $G_{-S}$ constitute the final clusters. A smaller value of $d$ leads to fewer singular points. A rule of thumb for its choice is $5\leqslant d \leqslant diam(G^k)$, where the diameter $diam(G^k)$ of the graph $G^k$ is the maximum separation of its vertices. The membership of each singular point $s$ may be assigned based on the membership of vertices in the neighbourhood of $s$. In particular, for each $s$ the path length to the solid clusters was computed, and its membership was assigned to the nearest cluster. In case of a tie, membership was assigned randomly. A \textsc{Mathematica} implementation is available\footnote{ \url{https://github.com/mariusz-tarnopolski/GraphBasedConnectivityClustering}}.

The choice of $(d,a)$ was determined by a grid search for all possibilities. First, the range of $d$ was restricted to $d\leqslant\lceil diam(G^k)/2 \rceil$ since for larger $d$ the set $V_v(d)$ extends over most of the data, and the resulting partitions contain just one cluster. Therefore, the scan was performed over $5\leqslant d\leqslant\lceil diam(G^k)/2 \rceil$, $a\leqslant 2d$. For each $d$, the $VI$ between $a$ and $a-1$ was computed and plotted against the associated $a$. For all data sets that were examined, most of the range of $a$ resulted in only two clusters and exhibited a long plateau in the values of $VI$. Among them, the minimal value was chosen, and the associated $a$ was recorded. For every $d$ an $a$ was therefore retrieved. The $VI$ was once again investigated, this time against $d$, whose value was chosen as the one with minimal $VI$. The obtained $(d,a)$ were then chosen as parameters for the final clustering.

\subsection{\texttt{fastdp}}
\label{sect::4.5}

The method of fast density peaks (\texttt{fastdp}; \citealt{sieranoja19}) also utilises the kNN graph. Its key concept is to identify cluster centres as regions with high local density $\rho$ and to be surrounded by more sparsely distributed points. The density at point $i$ is defined as the inverse of the mean distance from $i$ to other vertices of the kNN graph. The cluster centres should also be sufficiently separated from each other. The distance from point $i$ to the nearest point with a higher density is denoted by $\delta$; the latter is named a big brother (BB) of $i$. The number of clusters $K$ is a user-provided input parameter. Its choice can be made based on the decision plot of $(\rho,\delta)$: high values of $\gamma\equiv\rho\delta$ are likely to signalise clusters \citep{rodriguez14}. What constitutes a high value is a matter of context, though. The clusters are formed around $K$ points (density peaks) with the highest $\gamma$ values. The membership of the remaining points is assigned based on their BBs. The algorithm has a very fast and efficient $\textsc{C}$ implementation provided by the authors\footnote{\url{https://github.com/uef-machine-learning/fastdp}}.

\section{Results}
\label{results}

\subsection{Continuous $k$-nearest neighbour}

\begin{figure*}
\centering
\includegraphics[width=0.295\textwidth]{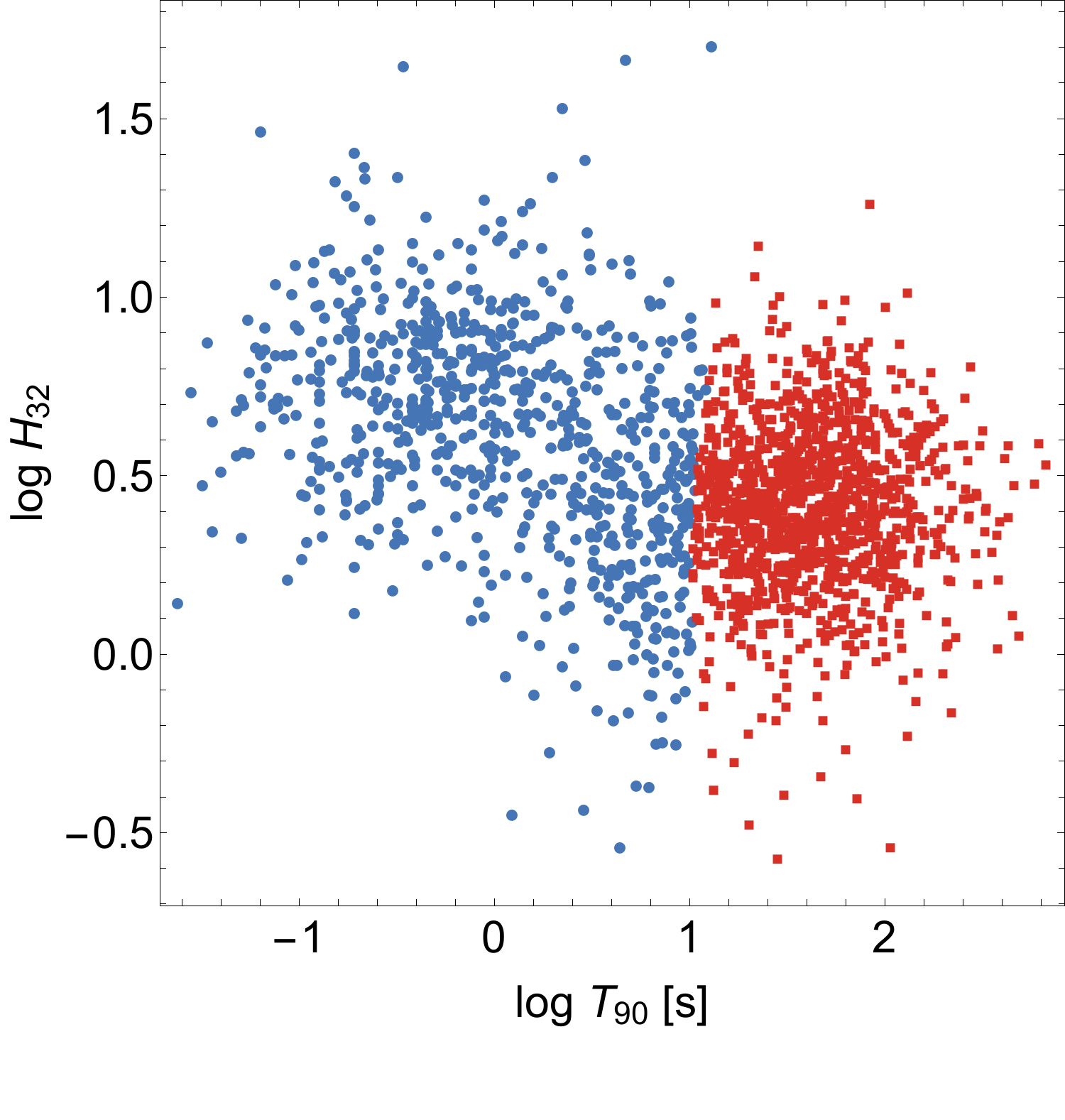}
\includegraphics[width=0.38\textwidth]{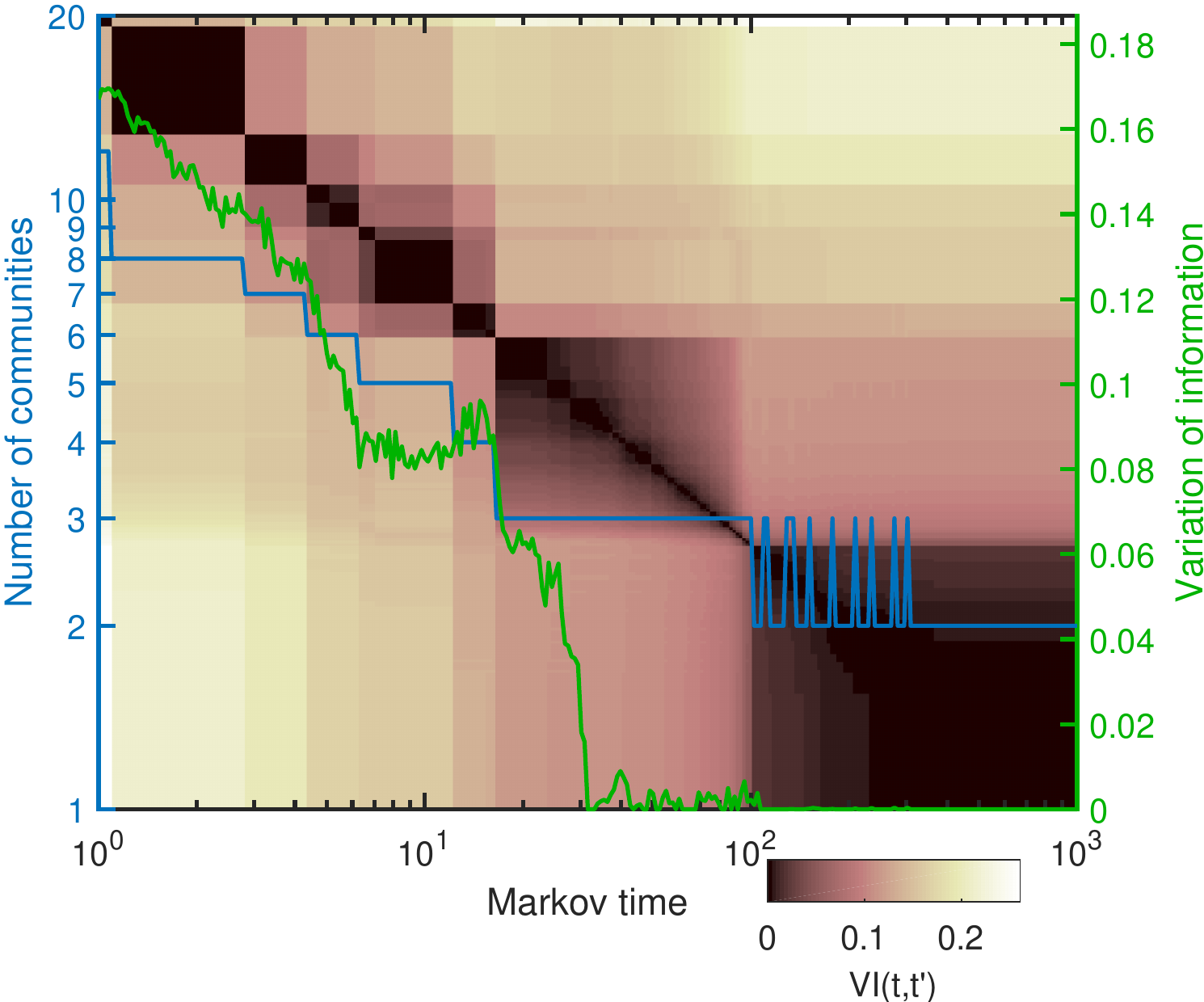}
\includegraphics[width=0.295\textwidth]{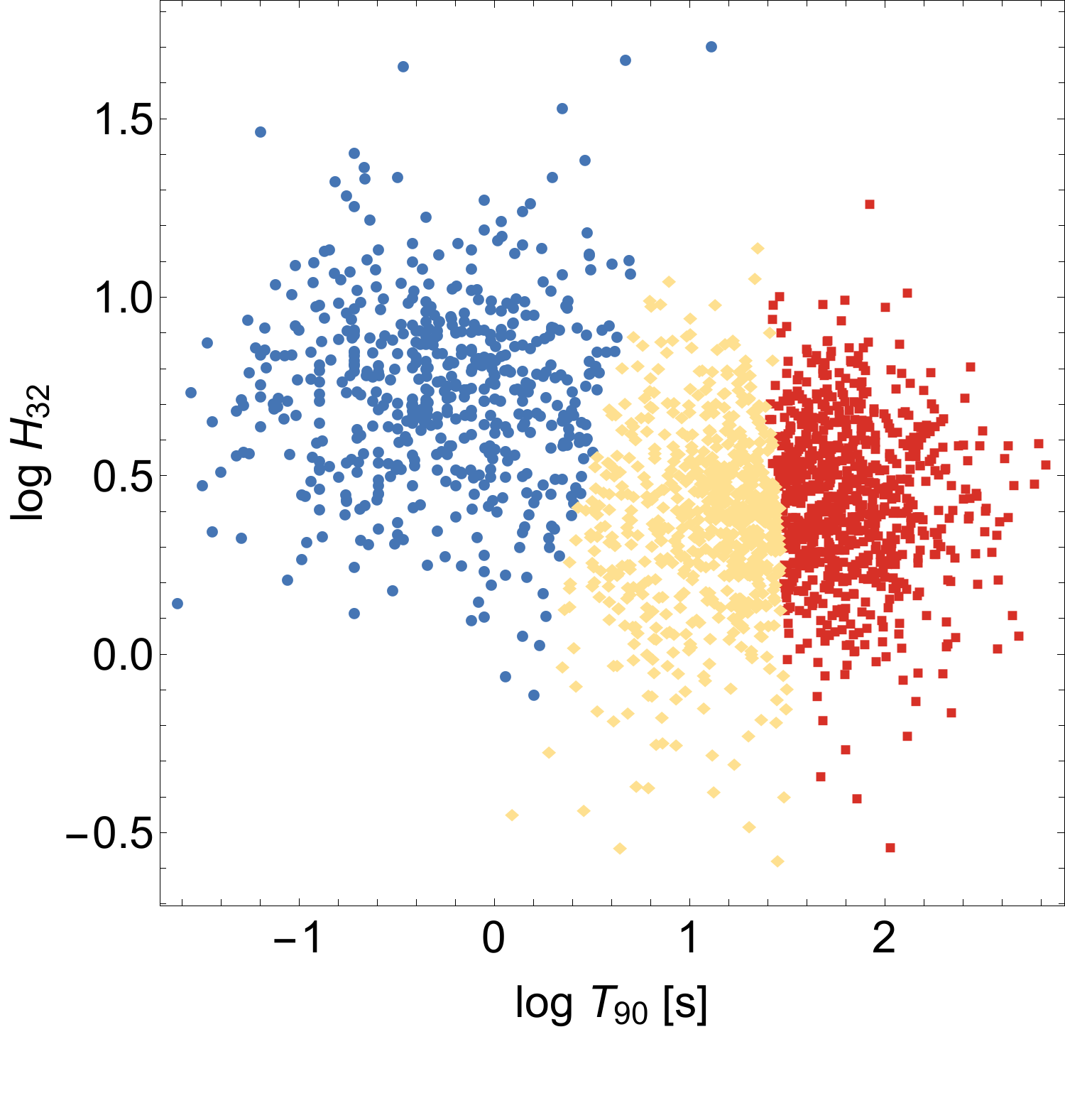}

\includegraphics[width=0.3\textwidth]{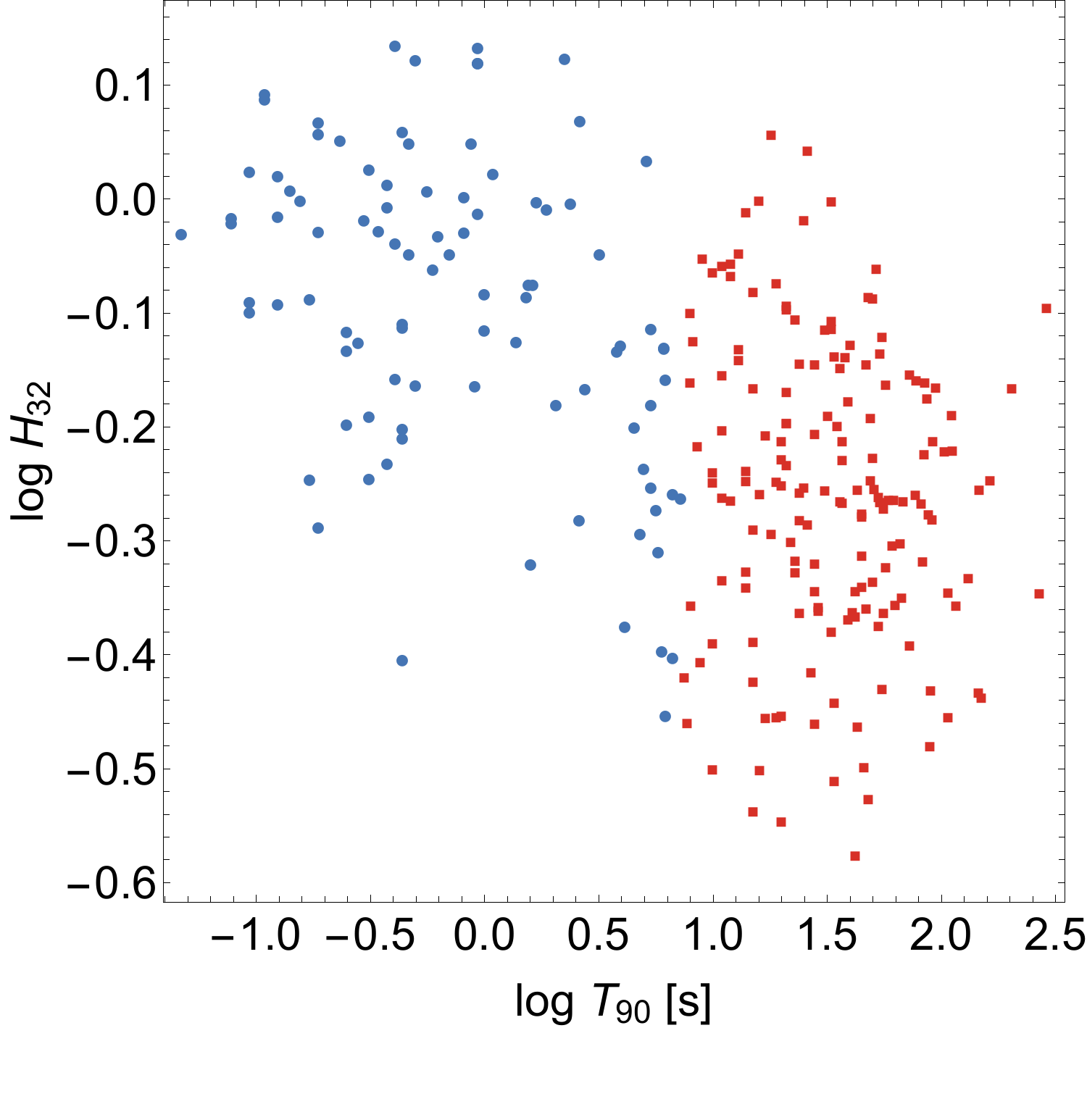}
\includegraphics[width=0.38\textwidth]{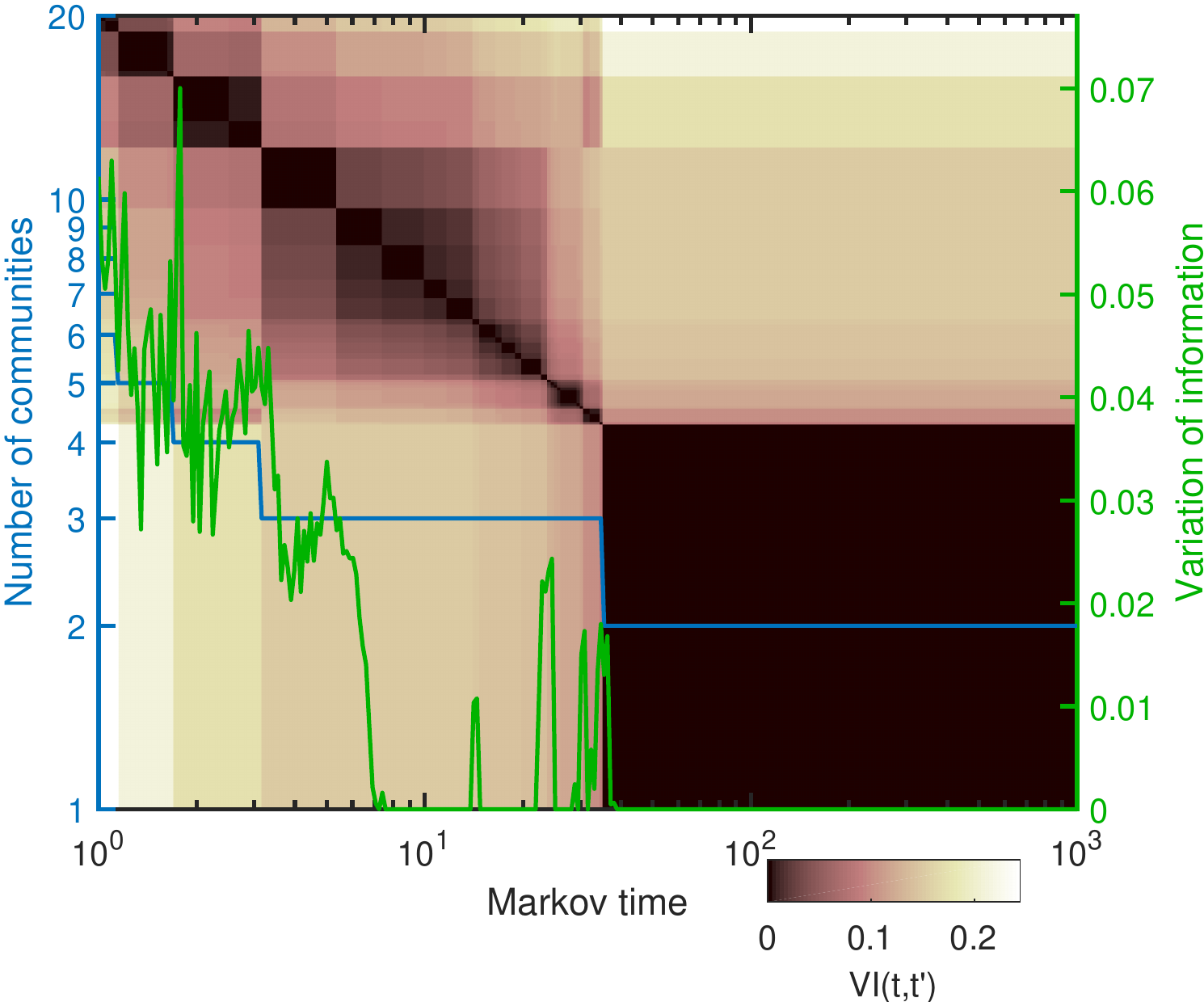}
\includegraphics[width=0.3\textwidth]{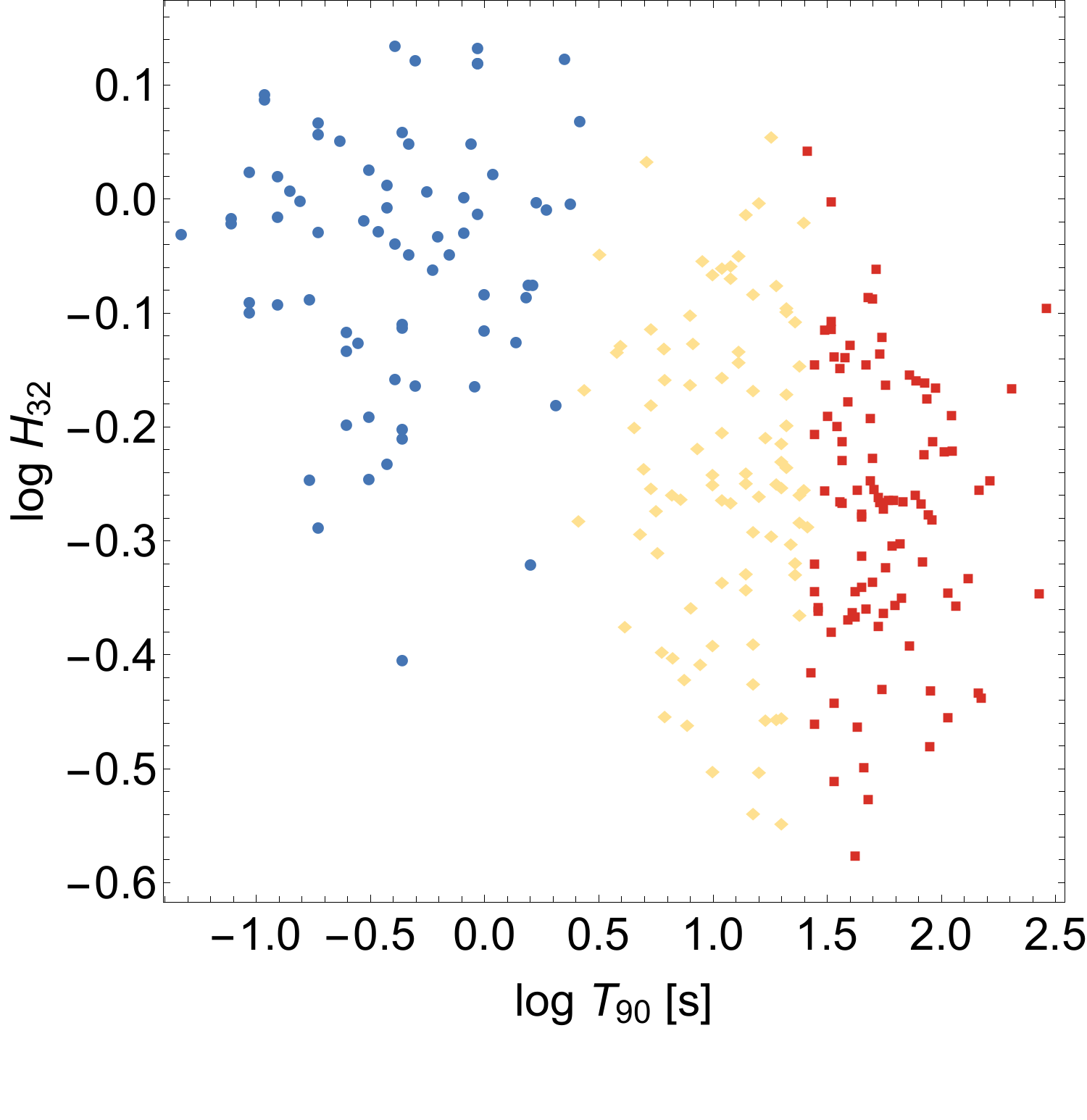}

\caption{BATSE (upper row) and Suzaku (bottom row) GRBs clustered with the CkNN algorithm. Left and right columns display partitioning into two and three groups, respectively; different colours (symbols -- circles, squares, and diamonds) symbolise communities. The middle column shows the evolution of the number of communities and $VI$ with Markov time $t$. The matrix plot in the background displays the $VI(t,t')$ matrix. Large blocks with small values on the diagonal relate to the number of communities (depicted with the blue line that) is dependent on the Markov time $t$. The sizes of the matrix blocks indicate the persistence of the corresponding number of communities. The $VI$ is depicted with the green line. Overall, a significant partition is characterised by a large block of small values of the $VI(t,t')$ matrix, and a low level of the $VI$.}
\label{fig2}
\end{figure*}

The CkNN algorithm was applied with $k=8$ and $\delta=2.4$, as recommended by \citet{liu20}, for Markov times $t\leqslant 10^3$, after which a saturation at two clusters was encountered. Two representative results are shown in Fig.~\ref{fig2}. In case of BATSE, the number of communities exhibits a prolonged plateau for $K=3$, during which the $VI$ drops to zero. This stage is followed by a prominent block corresponding to $K=2$. On the other hand, when Suzaku GRBs are considered, there is not much evidence for $K>2$, and hence only two clusters are detected. This suggests three communities for BATSE GRBs, and two for the Suzaku ones.

The pictures painted by the remaining four data sets (Fermi, Konus, Swift, and RHESSI) reveal a gradual transition of the domination of two over three groups, with the cases of Fermi and Konus resembling the results for BATSE, while Swift and RHESSI are more similar to Suzaku. This seems to be an effect of the sample size \citep{tarnopolski19d}, that is, in less numerous data sets there are simply not enough points to highlight more than two groups. Overall, the results point at either two or three clusters, hence do not conform with the works that find more GRB classes, for example five or seven \citep{acuner,ruffini18}.

\subsection{\texttt{CutPC}}

The \texttt{CutPC} algorithm contains only one free parameter, the tuning coefficient $\alpha$. While the authors \citep{li20} note that $\alpha=1$ should be suitable for most data sets, they observed that sometimes a better clustering is obtained when $\alpha$ is different. Therefore, for the considered GRB samples, the range $\alpha\in[0,2]$ was swept with a step $\Delta\alpha=0.01$. One seeks a stable partitioning, that is, a wide plateau of the number of communities in dependence on $\alpha$. Two very stable communities are observed for Suzaku, two or three communities for RHESSI (both partitions are very stable; the division into three groups is consistent with Gaussian mixture modelling done by \citealt{ripa09}), and formally four communities for Konus (but see further comments). Fermi and BATSE were inconclusive, that is, partitioning into more than one group does not yield stable clusters. This might be due to severe overlaps of the groups. The Swift data suffer from very non-uniform density of points: the long GRBs are confidently identified, but the short ones are either not classified as a cluster at all, or are randomly divided into several small classes.

\begin{figure}
\centering
\includegraphics[width=0.8\columnwidth]{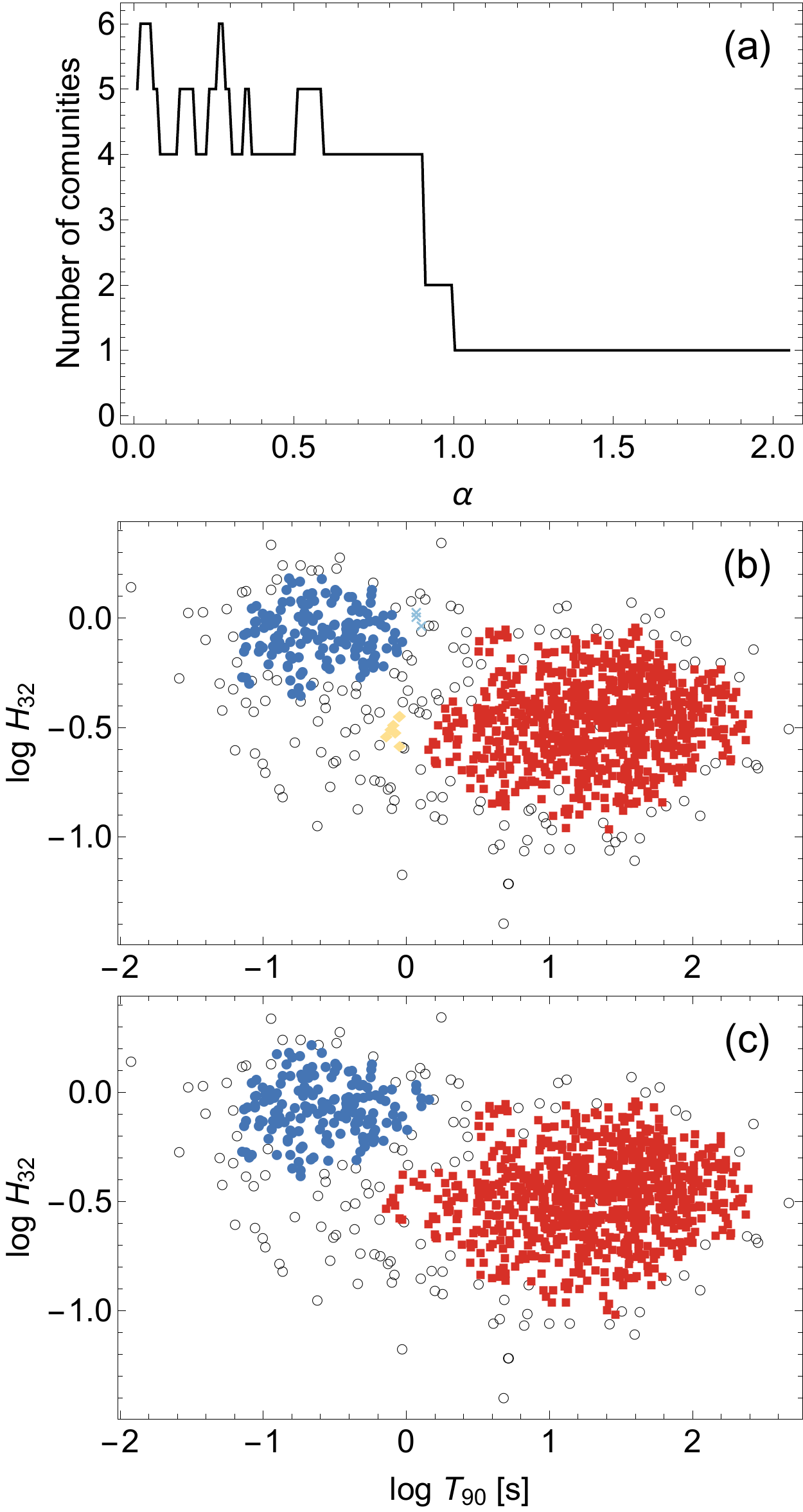}
\caption{Performance of the \texttt{CutPC} algorithm. (a) Dependence of the number of communities on $\alpha$ for the Konus data; and (b) clustering with $\alpha=0.70$ and (c) $\alpha=0.97$. Open black points are outliers, and the communities are denoted with different colours (symbols -- circles, squares, diamonds, and crosses).}
\label{fig3}
\end{figure}

\begin{figure}
\centering
\includegraphics[width=0.9\columnwidth]{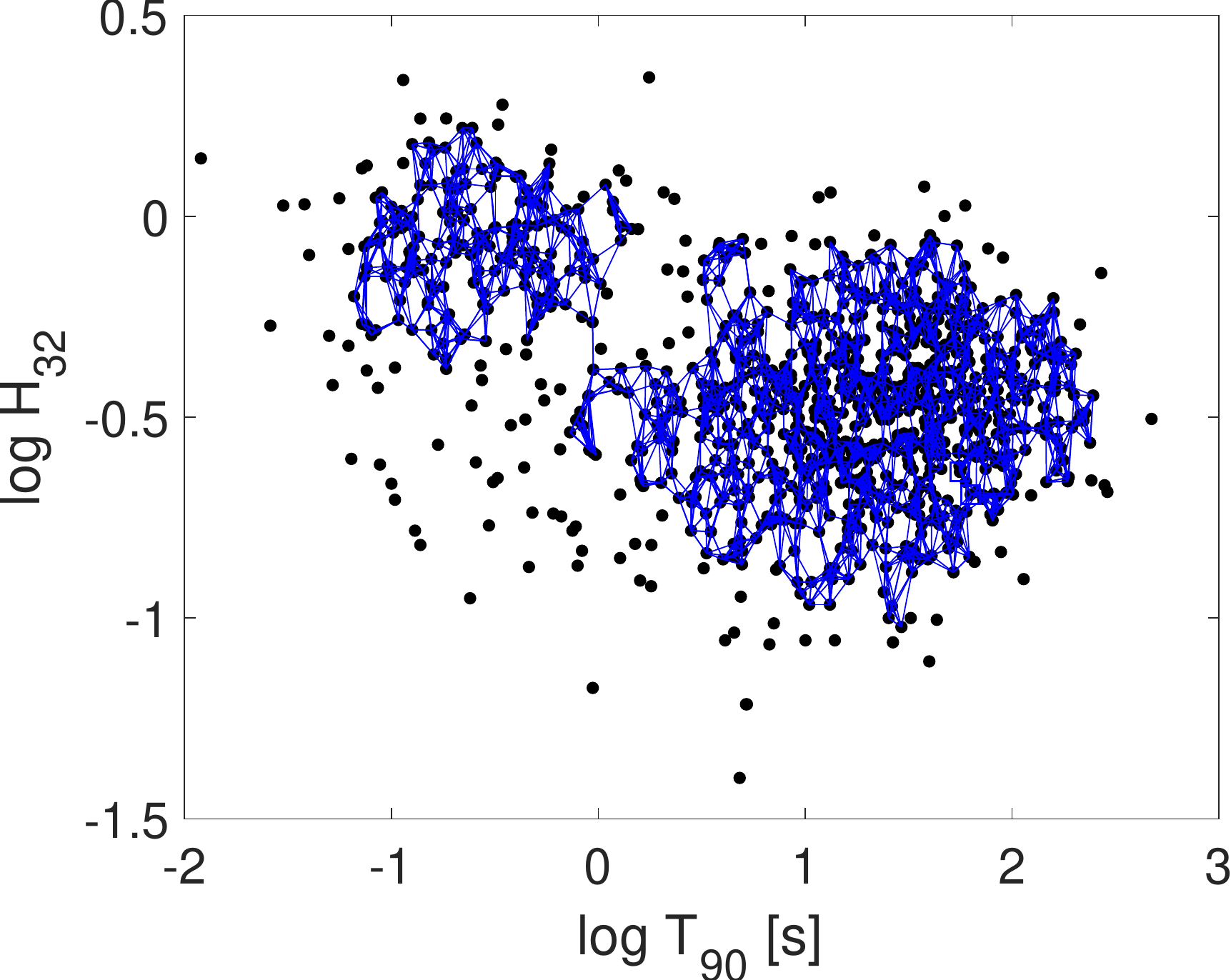}
\caption{Graph bridge that results with the \texttt{CutPC} algorithm applied to Konus data when $\alpha=1$ was employed, formally leading to only one cluster.}
\label{fig4}
\end{figure}

Furthermore, for example, in the case of Konus GRBs, clustering into four classes seems erroneous, as two big groups, easily attributed to short and long GRBs, are obtained, in addition to two small ones that contain only a few points each. Overall, many more points are classified by the algorithm as outliers than members of these groups. In Fig.~\ref{fig3} the dependence on $\alpha$ and the two partitions of Konus data are displayed. Figure~\ref{fig4} in turn demonstrates the final graph for $\alpha=1$. It is clear from this plot that there are two clusters, which are however connected by a bridge that leads to only one group, which is separated when $\alpha$ is slightly decreased to $\alpha=0.97$, leading to the two-group partitioning from Fig.~\ref{fig3}(c).

All in all, a two-group classification is easily visible, but there is no evidence to either support or reject the presence of more groups. The algorithm performs poorly on data with large variations in density, which is the case of GRB data. However, its future development aims to circumvent this problem \citep{li20}.

\subsection{Graph connectivity}

\begin{figure}
\centering
\includegraphics[width=0.9\columnwidth]{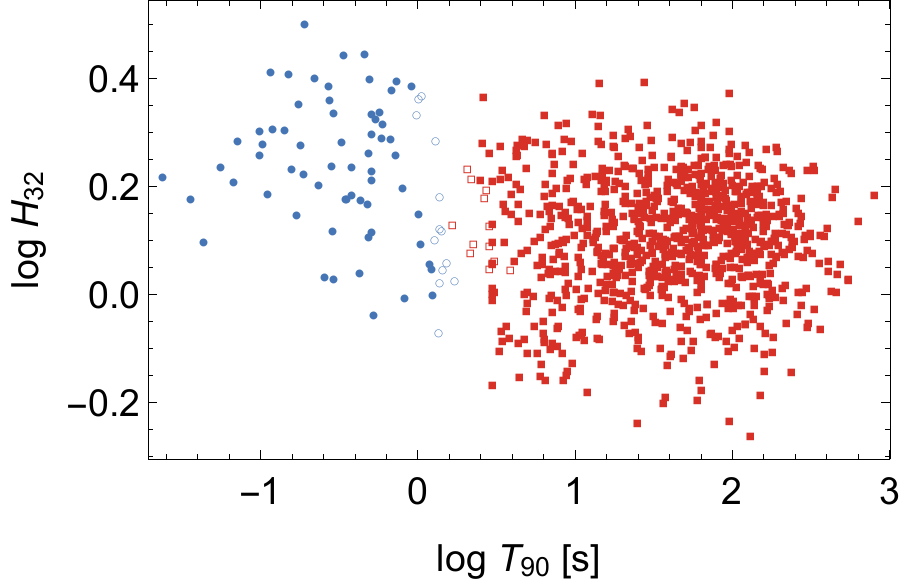}
\caption{Clusters in Swift GRBs identified by the connectivity method. Communities are marked with different colours (symbols -- circles and squares). Open symbols denote singular points, assigned to clusters based on proximity.}
\label{fig6}
\end{figure}

The connectivity-based method was employed with $k=6$ for all data sets. For Suzaku, RHESSI, Swift, and Konus, a stable partitioning was obtained for two clusters. They are consistent with both mixture modelling and other graph-based methods. In the case of Fermi and BATSE, however, the results are uninformative, that is, no plateau in $VI$ was reached before the number of communities dropped to one. A two-group partition of Swift GRBs is shown in Fig.~\ref{fig6}. 

\subsection{\texttt{fastdp}}

The \texttt{fastdp} algorithm was run for various values of $k$. For $2\leqslant k<10$, the partitions were very unstable, and they settled for $k\geqslant 10$. Eventually, $k=30$ was utilised for all data sets.

\begin{figure*}
\centering
\includegraphics[width=0.34\textwidth]{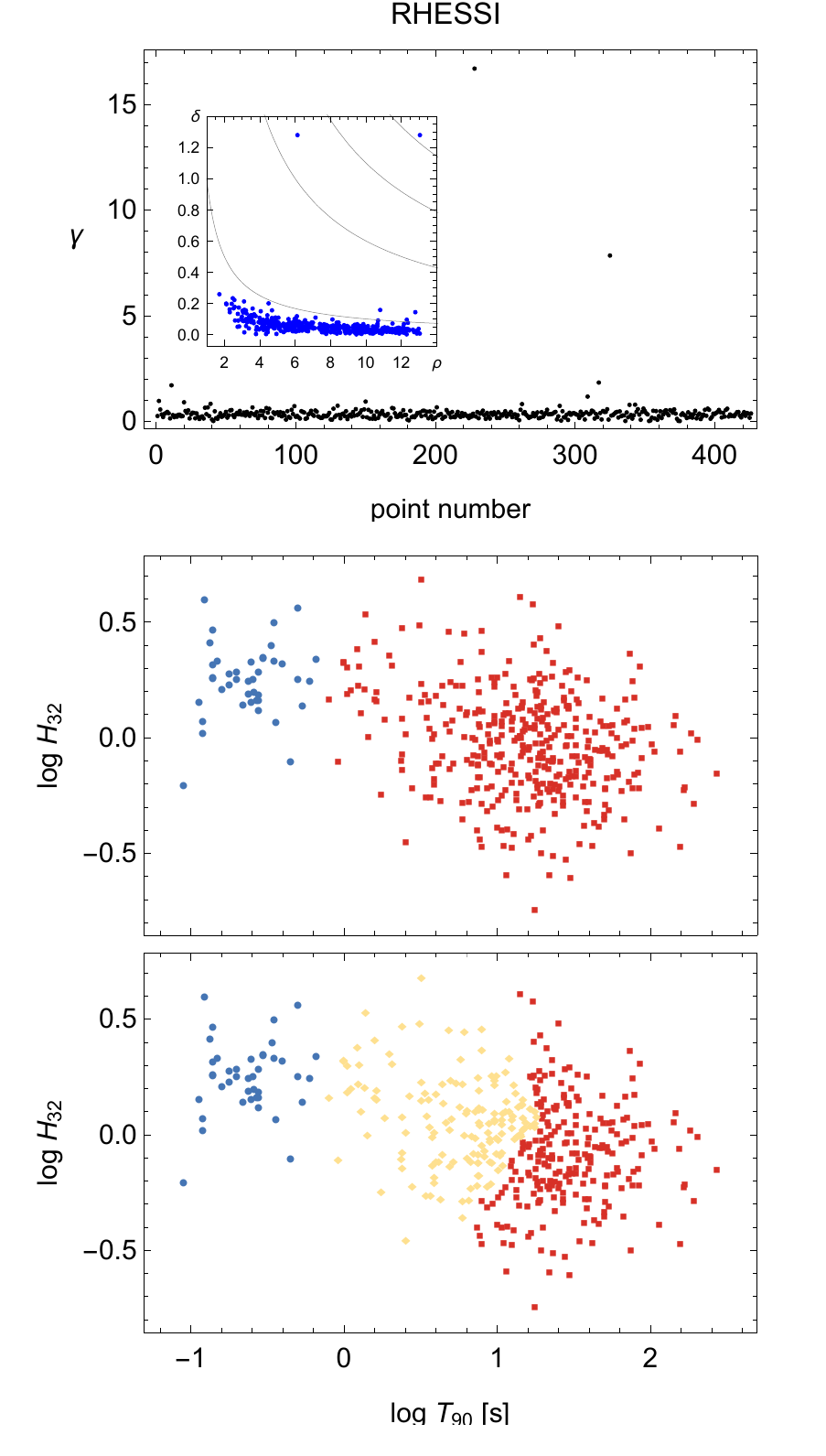}
\hspace{-0.5cm}
\includegraphics[width=0.34\textwidth]{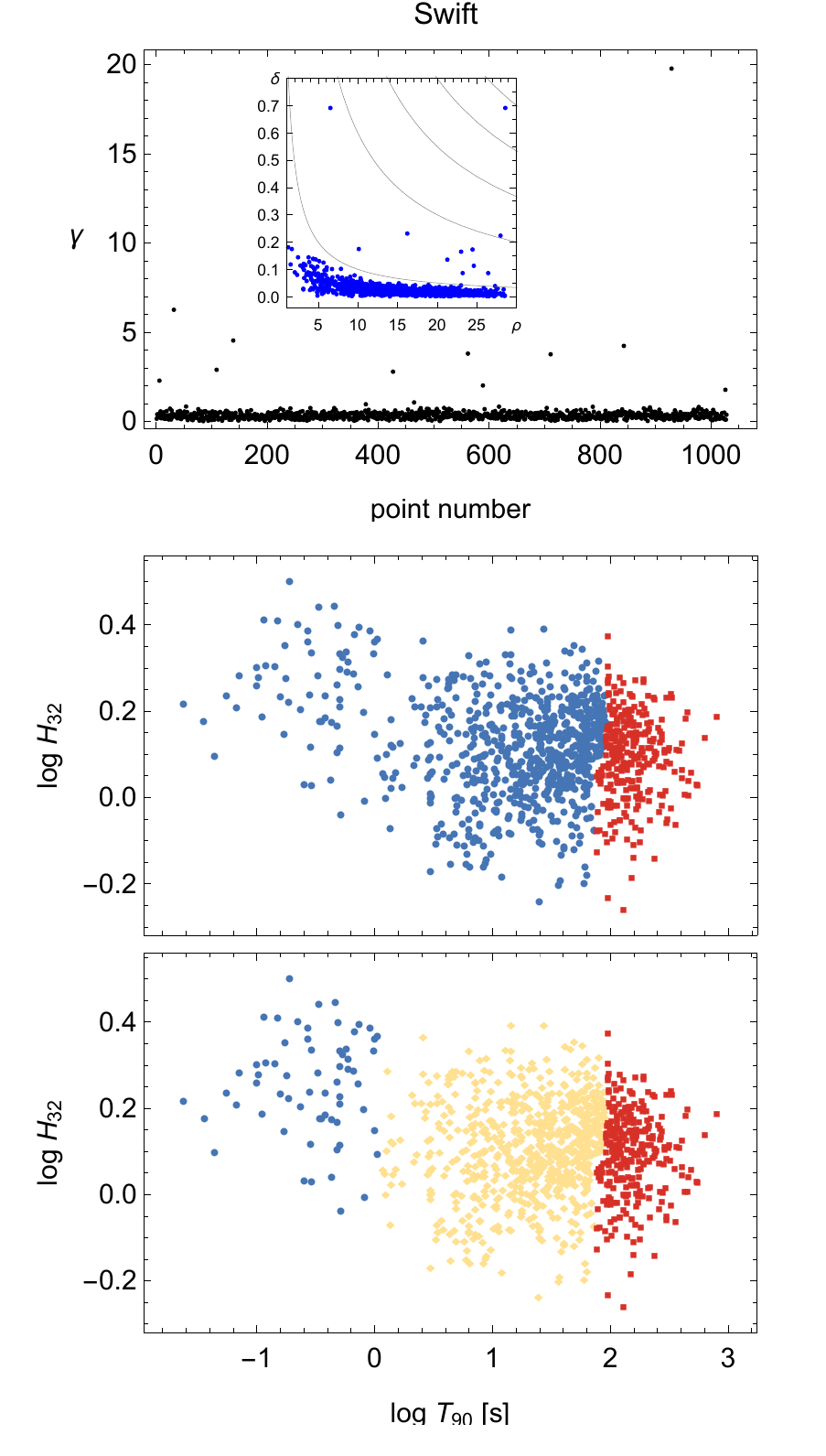}
\hspace{-0.5cm}
\includegraphics[width=0.347\textwidth]{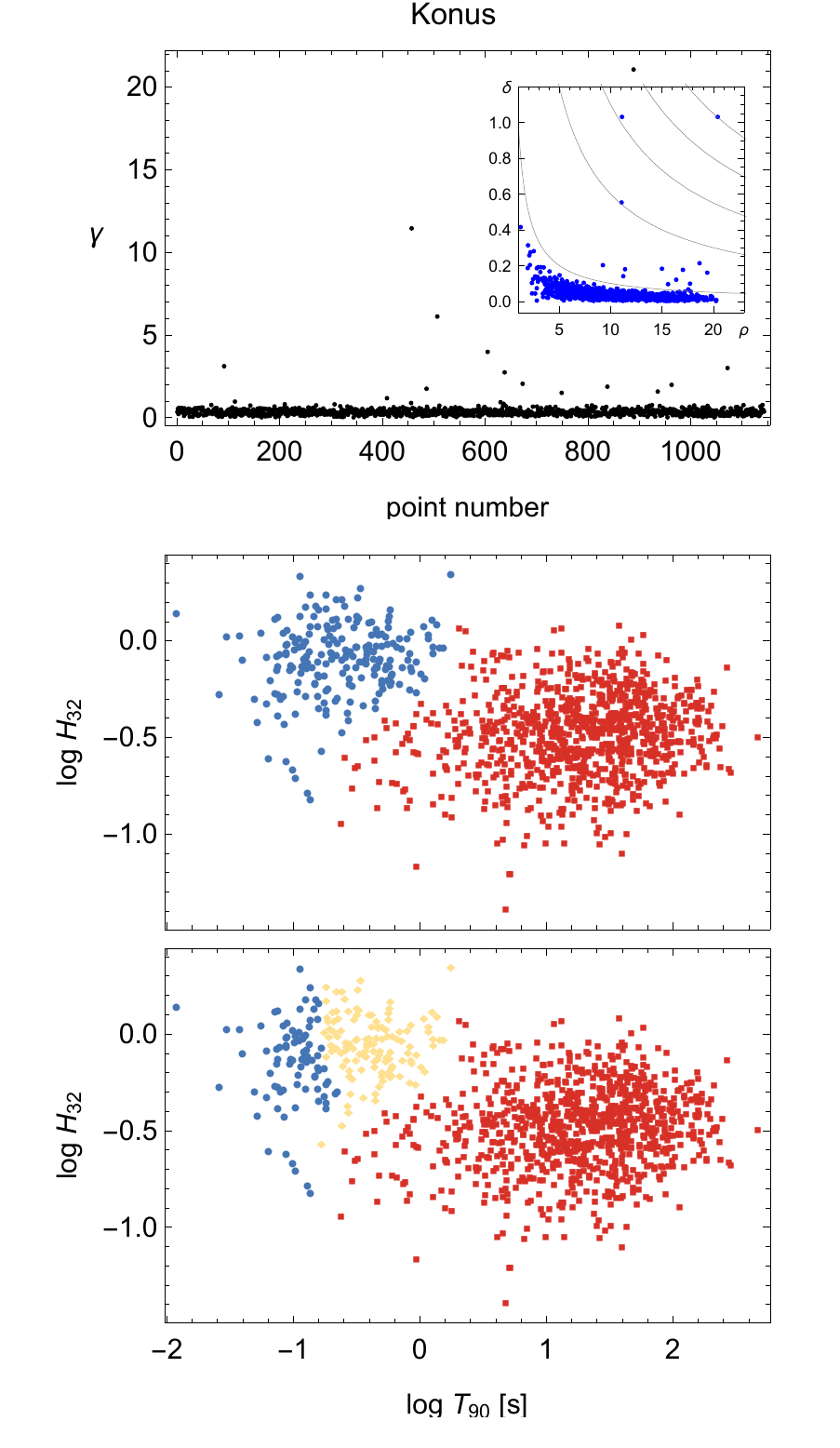}
\caption{Clusters identified in RHESSI (left column), Swift (middle column), and Konus GRBs (right column) with \texttt{fastdp}. The upper row shows the decision plots; the insets are the $(\rho,\delta)$ planes, with a few contours $\gamma \equiv \rho\delta = {\rm const.}$ displayed. The middle and bottom rows show the resulting partitions into two and three clusters, respectively, with different colours (symbols -- circles, squares, and diamonds) denoting the communities.}
\label{fig7}
\end{figure*}

For BATSE, the $\gamma$ values indicate two classes (Fig.~\ref{fig7}). However, there are several values clearly above the bulk around zero, but they are significantly lower than the two highest ones. A much less clear picture is obtained for Fermi, for which the $\gamma$ plot hints at about 15 classes, which seems highly unreliable and hence inconclusive. For Konus, it appears that clustering into three groups is appropriate; however, the division seems non-standard, yet similar to what was obtained with the Gaussian mixture model in the case of Fermi GRBs \citep{tarnopolski19a}. Swift's clustering into two groups is clearly erratic since the separation between short and long GRBs occurs at $T_{90}\lesssim 100\,{\rm s}$. Partitioning into three classes is more sensible; however, it is unclear why the long GRBs ought to be divided also at $T_{90}\lesssim 100\,{\rm s}$. It appears the algorithm just divides the cluster at the point of the highest local density into groups lying to the left and right of it. Finally, RHESSI and Suzaku are confidently divided into two standard groups, with only vague hints of a third one -- in such a case, the long class is cut into approximately equal halves.

\section{Discussion}
\label{discussion}

The objective of any clustering scheme is to divide a data set into groups within which points are most similar to each other as per some metric. There are various principles and heuristics for achieving this goal. In graph-based methods utilised herein, clusters were retrieved based on variants of the kNN graphs constructed from the samples at hand. These methods exploit not just the local densities, as the Gaussian mixture models do, but they explore the network relations (neighbours, big brothers, etc.) between pairs and subgroups of points, and how the structure of the graphs built from the samples (kNN graphs and its variants) depends on free parameters of the algorithms, seeking a robust, stable partition into communities (see Sect.~\ref{methods}). Graph-based algorithms can be hierarchical in nature, or can rely on distances between points or local densities (in the network sense) that signify centres of clusters. Usually they work well on groups sufficiently separated, but when overlaps become substantial -- as in case of GRBs -- the results become less unambiguous. The overlaps in the data seem to be the most serious source of biases. One needs to bear in mind there are also instrumental biases; for example, Swift is more sensitive in soft bands than BATSE was, therefore it yields a much lower fraction of short GRBs than other data sets. Hence the parallel investigation of six GRB samples was conducted herein to paint a more complete picture. For discussions on the instrumental effects and biases, readers can refer to \citet{shahmoradi13,shahmoradi15,ripa16,tarnopolski19a} and the references therein.

Some of the employed graph-based methods applied to GRBs worked relatively well for some data sets, while they gave rather unreliable results for others. Also, different algorithms gave different results for the same samples. It is thence worth emphasising that the mathematically formulated goals to be achieved by a particular algorithm can be apparently counter intuitive, especially when the desired outcome is not unequivocal. However, there are two significant outcomes to point out. The first one being that smaller samples (e.g. Suzaku) rather confidently exhibit two GRB classes, while bigger ones (e.g. BATSE) do not rule out the possibility of three being present. This observation that bigger samples are more liberal when it comes to accepting the possibility of a third class was also observed when using mixture models \citep{tarnopolski19d}. The second major result is that there are no hints of more than three groups (despite some works implying even five or seven).

Finally, a more robust classification could be achieved by combining the heuristics of different approaches, for example, weighted kNN graphs are likely to lead to better results. The weights, in turn, could be inferred from other GRB properties, not necessarily stemming from the metric relations of the considered parameter spaces. One should also note that all of the described algorithms can in principle operate on an arbitrary number of parameters, hence they do not need to be restricted to two-dimensional duration-hardness ratio spaces. This is a direction worth exploring in a subsequent work.

\section{Summary}
\label{summary}

The obtained GRB clusterings generally agree that there are two groups that can be associated with the conventional short--long dichotomy. In many cases, though, rejecting the possibility of a third group cannot be immediately discarded. In such cases, the resulting partitions split the long group in two parts that are roughly consistent with the intermediate and long GRB types. In summary, (i) it is therefore still unclear whether there are two or three distinct classes, but (ii) the graph-based methods led to no indications of, for example, five or seven groups, and (iii) after further development, such methods are promising tools for looking at GRBs from a different perspective, and they can prove to be useful in classifying other astronomical objects, such as stars or galaxies as well.

\begin{acknowledgements}
The author acknowledges support by the Polish National Science Centre through the OPUS grant No. 2017/25/B/ST9/01208.
\end{acknowledgements}

\bibliography{bibliography}{}

\end{document}